\documentclass{ifacconf}

\usepackage{graphicx}      
\usepackage{amssymb}
\usepackage{amsmath}
\usepackage{mathtools}
\usepackage{natbib}        
\usepackage{tikz}               
\usepackage{pgfplots}
\usepackage{color}
\allowdisplaybreaks

\colorlet{istorange}{orange}
\colorlet{istgreen}{green!50!black}
\colorlet{istblue}{blue} 
\colorlet{istred}{red!90!black}

\newtheorem{ass}{Assumption}
\newtheorem{lemma}{Lemma}

\newtheorem{remark}{Remark}
\newcommand*{\QEDB}{\hfill\ensuremath{\square}}%

\begin{document}
\begin{frontmatter} 

\title{Saving Tokens in Rollout Control with Token Bucket Specification} 

\author[FJ]{Florian Jaumann} 
\author[WA]{Stefan Wildhagen} 
\author[WA]{Frank Allg\"ower}

\address[FJ]{University of Stuttgart, Germany (st115660@stud.uni-stuttgart.de).}
\address[WA]{Institute for Systems Theory and Automatic Control, University of Stuttgart, Germany (\{wildhagen,allgower\}@ist.uni-stuttgart.de). }

\thanks[footnoteinfo]{Funded by the Deutsche Forschungsgemeinschaft (DFG, German Research Foundation) - 285825138; 390740016.}

\begin{abstract}
We consider a communication network over which transmissions must fulfill the so-called token bucket traffic specification, with a rollout (i.e., predictive) controller that both schedules transmissions and computes the corresponding control values. In the token bucket specification, a transmission is allowed if the current level of tokens is above a certain threshold. Recently, it has been shown that having a full bucket at the time of a set point change significantly improves the control performance as compared to when the bucket level is low. In this work, we develop mechanisms that guarantee that the bucket fills up after the controlled plant has converged to a set point. To do this, we consider two different setups. First, we consider that all transmissions over the network must fulfill the token bucket specification and show convergence to the upper sector of the bucket by adding a slight terminal cost on the bucket level. Afterwards, we consider a modified network which additionally features a direct link over which transmissions need not fulfill the token bucket specification. In this setup, we prove convergence of the bucket level exactly to the upper rim. These mechanisms enable a similar level of flexibility as event-triggered control: In converged state, little communication is used while in precarious operating conditions, a burst of transmissions is possible. Other than event-triggered approaches, the proposed methods allow to specify the network traffic beforehand by means of the token bucket. Lastly, we validate the proposed approaches in a numerical example.
\end{abstract}


\end{frontmatter}

\section{Introduction}
Networked Control Systems (NCS), where information is transmitted over a communication network, offer many benefits over traditional point-to-point wired communication links such as greater flexibility, lower cost and simpler maintenance. Then again, NCS induce imperfections like transmission delays, quantization errors and packet loss which can degrade the control performance. Since such imperfections especially occur if the network is congested, a proper tradeoff between communication effort and control performance is vital. A number of different approaches that address this issue were considered in the literature.

Two popular methods in this direction are event-triggered control and self-triggered control (see \cite{6425820} for an overview). In event-triggered control, a triggering condition is persistently monitored and if it is violated, a transmission is triggered. In self-triggered control, in contrast, the controller acts proactively and determines the next transmission time in advance based on predictions from recent data. In both approaches, transmissions are reduced as much as possible while still guaranteeing stability or a certain performance level, however, a traffic specification is typically not considered. Another way to address the performance-communication tradeoff is to consider certain prespecified traffic constraints, so-called traffic specifications (TSs), directly in the controller. This is done in rollout approaches in NCS, which are essentially a special form of model predictive control (MPC). In a receding horizon fashion, both the control inputs and transmission decisions are determined on the basis of a cost functional. Transmissions are only scheduled if allowed by the TS. Such an approach was taken in \cite{6315640}, \cite{6882787} and \cite{article}, where a window-based TS was used, allowing a certain number of transmissions in each disjoint window.

In \cite{8550568}, the token bucket TS was used to characterize the traffic, in which a transmission is allowed if the current number of tokens in a bucket is at least as high as the cost of a transmission. Tokens are added to the bucket at a constant rate. The benefit of such a dynamic TS lies in the possibility to build up tokens in situations where little communication is needed, and using them at a later point, e.g., in the form of a quick burst of transmissions when a set point change occurs. In \cite{8723153}, a rollout approach in combination with the token bucket specification was investigated. It was shown that a full bucket can greatly improve the control performance, however, no mechanism to fill up the bucket was provided therein. In this work, we want to achieve a refilling bucket by renouncing transmissions in situations where they are not necessary, i.e., when the plant has converged. A related idea was presented for the case of disturbances on the plant and a model-based actuator in \cite{Antunes2016ConsistentEM} under the term consistent event-triggered control: An efficient transmission scheme for NCS should outperform time-triggered control and save communication resources in uncritical situations, i.e., if no disturbance is acting.

In this work, we contribute mechanisms to save tokens in order to attain a full bucket after the plant has converged. First, we consider the NCS setup with the token bucket TS introduced in \cite{8723153} and prove that the bucket level converges to the upper sector of the bucket when an arbitrarily small terminal cost on the bucket level is added in the MPC optimization problem. Then, we extend the network by an additional direct link, over which transmissions do not consume any tokens. For this setup, we can prove convergence of the bucket level exactly to the upper rim.

With these mechanisms, rollout control may achieve a similar level of flexibility as event- and self-triggered approaches, in the sense that there is little communication in converged state, and the possibility to use quick bursts of transmissions in precarious operating conditions such as a set point change or when a disturbance is acting. In addition, the token bucket enforces a certain TS, whereas no such guarantee can be given in event-triggered control.

The remainder of this paper is structured as follows. In Section \ref{sec:problem_setup}, the considered NCS setup and the control scheme is introduced, while in Section \ref{sec:extended_term}, convergence of the bucket level to the upper sector is proven. In Section \ref{sec:periodic}, a modified NCS setup is introduced and convergence of the bucket level to the upper rim in this modified setup is established. In Section \ref{sec:num_example}, a numerical example is provided and in Section \ref{sec:conclusion}, a summary and an outlook are given.

Let $\mathbb{I}$ and $\mathbb{R}$ denote the set of all integers and real numbers, respectively. We denote $\mathbb{I}_{[a,b]}\coloneqq\mathbb{I}\cap[a,b]$ and $\mathbb{I}_{\ge a}\coloneqq\mathbb{I}\cap[a,\infty)$, $a,b\in\mathbb{I}$. We denote by $A > 0$ a positive definite matrix $A \in \mathbb{R}^{n\times n}$. For the weighted vector norm with $v \in \mathbb{R}^n$ and $A \in \mathbb{R}^{n \times n}, A > 0$, we write $\Vert v \Vert^2_A\coloneqq v^T A v$.
\section{Problem Setup and Proposed Mechanism}
\label{sec:problem_setup}
\subsection{Plant and General NCS Setup}
\begin{figure}[!h]
	\centering
	\resizebox {0.5\textwidth} {!} {
		\tikzset{every picture/.style={line width=0.75pt}} 

\begin{tikzpicture}[x=0.75pt,y=0.75pt,yscale=-0.9,xscale=1]

\draw   (90,48) .. controls (90,43.58) and (93.58,40) .. (98,40) -- (162,40) .. controls (166.42,40) and (170,43.58) .. (170,48) -- (170,72) .. controls (170,76.42) and (166.42,80) .. (162,80) -- (98,80) .. controls (93.58,80) and (90,76.42) .. (90,72) -- cycle ;
\draw   (240,48) .. controls (240,43.58) and (243.58,40) .. (248,40) -- (282,40) .. controls (286.42,40) and (290,43.58) .. (290,48) -- (290,72) .. controls (290,76.42) and (286.42,80) .. (282,80) -- (248,80) .. controls (243.58,80) and (240,76.42) .. (240,72) -- cycle ;
\draw   (360,48) .. controls (360,43.58) and (363.58,40) .. (368,40) -- (442,40) .. controls (446.42,40) and (450,43.58) .. (450,48) -- (450,72) .. controls (450,76.42) and (446.42,80) .. (442,80) -- (368,80) .. controls (363.58,80) and (360,76.42) .. (360,72) -- cycle ;
\draw   (190.5,138) .. controls (190.5,133.58) and (194.08,130) .. (198.5,130) -- (352,130) .. controls (356.42,130) and (360,133.58) .. (360,138) -- (360,162) .. controls (360,166.42) and (356.42,170) .. (352,170) -- (198.5,170) .. controls (194.08,170) and (190.5,166.42) .. (190.5,162) -- cycle ;
\draw    (88,60.04) -- (39.5,61) ;

\draw [shift={(90,60)}, rotate = 178.87] [fill={rgb, 255:red, 0; green, 0; blue, 0 }  ][line width=0.75]  [draw opacity=0] (8.93,-4.29) -- (0,0) -- (8.93,4.29) -- cycle    ;
\draw    (40,150) -- (39.5,61) ;

\draw    (40,150) -- (190,150) ;

\draw    (238,60) -- (170,60) ;

\draw [shift={(240,60)}, rotate = 180] [fill={rgb, 255:red, 0; green, 0; blue, 0 }  ][line width=0.75]  [draw opacity=0] (8.93,-4.29) -- (0,0) -- (8.93,4.29) -- cycle    ;
\draw    (358,60) -- (290,60) ;

\draw [shift={(360,60)}, rotate = 180] [fill={rgb, 255:red, 0; green, 0; blue, 0 }  ][line width=0.75]  [draw opacity=0] (8.93,-4.29) -- (0,0) -- (8.93,4.29) -- cycle    ;
\draw    (360,150) -- (490,150) ;

\draw    (450,60) -- (490,60) ;

\draw    (490,60) -- (490,80) ;

\draw    (490,80) -- (496.5,87) ;

\draw    (490,90) -- (490,150) ;

\draw  [dash pattern={on 0.84pt off 2.51pt}] (340,30) -- (530,30) -- (530,110) -- (340,110) -- cycle ;

\draw  [dash pattern={on 0.84pt off 2.51pt}] (80,120) -- (400,120) -- (400,180) -- (80,180) -- cycle ;

\draw (130,60) node  [align=left] {Actuator};
\draw (265,60) node  [align=left] {Plant};
\draw (405,60) node  [align=left] {Controller};
\draw (275.25,150) node  [align=left] {Token Bucket TS};
\draw (204,51) node  [align=left] {$u_p(k)$};
\draw (319.5,51) node  [align=left] {$x_p(k)$};
\draw (480,51) node  [align=left] {$u_c(k)$};
\draw (510,91) node  [align=left] {$\gamma(k)$};
\draw (385,100) node  [align=left] {Smart sensor};
\draw (110,169) node  [align=left] {Network};

\end{tikzpicture}
	}
	\caption{NCS configuration with token bucket TS.}
	\label{fig:NCS_config}
\end{figure}
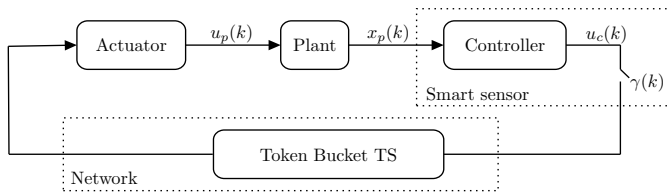
We consider the NCS configuration shown in Fig. \ref{fig:NCS_config}. It comprises a nonlinear discrete-time plant
\begin{equation}
x_p(k+1)=f_p(x_p(k),u_p(k))
\label{eq:plant}
\end{equation}
with plant state $x_p(k) \in \mathbb{X}_p \subseteq\mathbb{R}^{n_p}$ and input $u_p(k) \in \mathbb{U}_p \subseteq\mathbb{R}^{m_p}$ at time $k \in \mathbb{I}_{\geq 0}$, and $f_p(0,0)=0$. Both $\mathbb{X}_p$ and $\mathbb{U}_p$ are assumed to be closed and to contain the origin. In order to measure performance, the quadratic cost 
\begin{equation}
\Vert x_p \Vert_Q^2+\Vert u_p \Vert_R^2 \quad \text{with} \quad Q,R > 0
\label{eq:gen_cost}
\end{equation} 
is associated with the plant \eqref{eq:plant}. The full state of the plant is sensed at the smart sensor. Based on solving an optimization problem, the smart sensor decides both the control input sequence $u_c(\cdot)$ and the binary transmission sequence $\gamma(\cdot)$. The binary sequence indicates when information should be sent over the network.  In case of a transmission, $\gamma(k)=1$ holds and the control input is transmitted through the network to the actuator. The actuator, which has no computational capabilities, no clock or memory, merely holds the last received input. If a new input is received, it is immediately applied to the plant. If no transmission over the network is triggered ($\gamma(k)=0$), the actuator holds the old input such that the zero-order hold dynamics take the form 
\begin{equation*}
u_p(k)=\gamma(k)u_c(k)+(1-\gamma(k))u_p(k-1)
\end{equation*}
with initial condition $u_p(-1) \in \mathbb{U}_p$. The sensor keeps track of the previously applied input $u_p$ in the form
\begin{equation}
u_s(k)=u_p(k-1).
\nonumber
\end{equation}

It is assumed that the network traffic must fulfill the so-called token bucket TS. Despite being commonly applied in computer networks (see \cite{tanenbaum2010computer}), it was first used in the control context in \cite{8550568}. For a detailed description of the token bucket model, we refer the reader to \cite{8723153}. Here, only a short outline shall be given. The token bucket TS builds on an analogy to a bucket which stores a certain amount of tokens $\beta(k)$ at time $k$. New tokens are added to the bucket with a constant rate of $g \in \mathbb{I}_{\geq 1}$, while the cost of a transmission is $c \in \mathbb{I}_{\geq g}$. If the bucket size $b \in \mathbb{I}_{\geq c}$ is reached, further arriving tokens are discarded. This results in the dynamics
\begin{equation}
\beta(k+1)=\min \lbrace \beta(k)+g-\gamma(k)c,b\rbrace.
\label{eq:beta_dyn}
\end{equation}
The token bucket TS is fulfilled if the bucket is not drained, which translates to the constraint $\beta(k)\in \mathbb{I}_{[0,b]}$. It typically holds that $c$ is larger than $g$, such that it is not possible to transmit at any time step. However, according to the above model, a transmission is still guaranteed to be possible at least every $q\coloneqq \lceil \frac{c}{g}\rceil$ time instances. In effect, the token bucket TS enforces that the average transmission rate of the control application is at most $\frac{g}{c}$.
 
Combining the token bucket TS with the other components of the NCS results in the overall state $x\coloneqq [x_p^T,u_s^T,\beta]^T$ and control $u\coloneqq [u_c^T,\gamma]^T$. The overall dynamics are characterized by
  \begin{equation}
  x(k+1)=f(x(k),u(k)) 
  \label{eq:dyn_all}
   \end{equation}
  \begin{equation*}
  \begin{split}
  \text{with} \quad 
  f(x,u)\coloneqq \begin{bmatrix}
  f_p(x_p,\gamma u_c+(1-\gamma)u_s) \\\gamma u_c+(1-\gamma)u_s\\\min\lbrace \beta + g - \gamma c,b \rbrace
  \end{bmatrix}
  \end{split}
 \label{eq:dyn_matrix}
 \end{equation*}
 with state $x(k)\in \mathbb{X}_p \times \mathbb{U}_p \times \mathbb{I}_{[0,b]} \eqqcolon \mathbb{X}\subseteq \mathbb{R}^{n_p+m_p+1}$ and input $u(k)\in \mathbb{U}_p \times \lbrace 0,1\rbrace \eqqcolon \mathbb{U} \subseteq \mathbb{R}^{m_p+1}$. The cost \eqref{eq:gen_cost} associated with the plant $\ell:\mathbb{X}\rightarrow\mathbb{R}$ becomes
 \begin{equation*}
 \ell(x,u)=\Vert x_p \Vert_Q^2 + \gamma\Vert u_c  \Vert_R^2 + (1-\gamma)\Vert u_s  \Vert_R^2.
 \label{eq:stage_cost}
 \end{equation*}
 
\subsection{Rollout Control Scheme}
\label{sec:control_scheme}	
 The rollout control scheme described in \cite{8723153} is applied in this paper. It reads as follows:
 \begin{itemize}
 	\item[1)]{ At time $k=jM, j \in \mathbb{I}_{\geq 0}$, measure $x(jM)$ and solve the optimization problem $\mathbb{P}(x(jM))$ defined by
 		\begin{align*}
 		\begin{split}
 		&\begin{aligned}
 		V^*(x(jM))\coloneqq\min\limits_{u(\cdot \vert jM)}\sum\limits_{i=0}^{N-1}&\ell(x(i \vert jM),u(i \vert jM))\\&+V_f(x(N\vert jM))
 		\end{aligned}  	\\		
 		&\begin{aligned}
 		\text{s.t.} \quad  &x(i+1\vert jM)=f(x(i\vert jM),u(i\vert jM))\\
 		&x(i\vert jM) \in \mathbb{X}, \quad u(i\vert jM) \in \mathbb{U}, \quad \forall i \in \mathbb{I}_{[0,N-1]} \\
 		&x(0\vert jM) = x(jM), \quad x(N\vert jM) \in \mathbb{X}_f
 		\end{aligned}
 		\end{split}
 		\end{align*}
 		and denote its optimizer by $u^*(\cdot\vert jM)$.
 	} 
 	\item[2)]{Apply $u(jM+i)=u^*(i\vert jM)$ for $i \in \mathbb{I}_{[0,M-1]}$.}
 	\item[3)]{Set $j \leftarrow j +1$ and go to 1).}
 	
 \end{itemize}
According to the control scheme, the controller is activated every $M\coloneqq rq$, $r \in \mathbb{I}_{\geq 1}$ time steps, where $r$ is user-chosen. It minimizes the stage cost $\ell$ and the terminal cost $V_f: \mathbb{X} \rightarrow \mathbb{R}$ along all feasible trajectories of \eqref{eq:dyn_all} over the prediction horizon $N \in \mathbb{I}_{\geq M}$. A terminal cost together with the terminal constraint $x(N\vert jM) \in \mathbb{X}_f \subseteq \mathbb{X}$ is added to the costs and constraints posed by the NCS to ensure recursive feasibility and convergence, supposed that  $\mathbb{P}$ is feasible at initial time.
\subsection{Proposed Mechanism}
In \cite{8723153}, a terminal cost $V_f(x)=V_{f,p}(x_p)$, $V_{f,p}:\mathbb{X}_p \rightarrow \mathbb{R}$, only penalizing the plant state $x_p$ was used to guarantee convergence of $x_p$ and $u_s$. Our aim in this paper, however, is to guarantee also convergence of the bucket level to the upper rim. To do this, we extend the previously used terminal cost by a term on the bucket level
\begin{equation}
V_f(x) \coloneqq V_{f,p}(x_p) + V_{f,\beta}(\beta), 
\label{eq:term_cost_total}
\end{equation}
where $V_{f,\beta}: \mathbb{I}_{[0,b]}\rightarrow \mathbb{R}$ denotes the terminal cost on the bucket level $\beta$. We consider a quadratic $ V_{f,\beta}$ of the form
\begin{equation}
V_{f,\beta}(\beta) =\sigma (b^2-\beta^2)
\label{eq:beta_term_cost}
\end{equation}
with scaling parameter $\sigma \in (0,\infty)$. Hence, the terminal cost on the bucket level is highest  when the bucket is empty. When the bucket is full, $V_{f,\beta}(\beta)=0$ holds.

\begin{remark}
Other forms of $V_{f,\beta}$ are also possible, as long as they are positive definite with respect to $\beta=b$. Then, similar guarantees as shown later could be derived. The particular cost \eqref{eq:beta_term_cost} is chosen to facilitate calculations later.
\end{remark}

The parameter $\sigma$ should be chosen such that $V_{f,\beta}$ is very small compared to $\ell$ and $V_{f,p}$ to achieve the following in closed loop. Our primary objective is to achieve a high performance with respect to $\ell$ and $V_{f,p}$, such that $x_p$ and $u_s$ are quickly brought to equilibrium by the controller. In this phase, a slight terminal cost on the bucket level has little effect on the closed-loop behavior. Our secondary objective, after $x_p$ and $u_s$ have converged, is to minimize $V_{f,\beta}$, such that transmissions are saved and the bucket refills. Achieving such a behavior is our main motivation to add a slight terminal cost on the bucket level. 
\section{Guaranteeing Convergence of the Bucket Level to the Upper Sector }
\label{sec:extended_term}
\subsection{Preliminaries}
In this section, we consider the dynamics \eqref{eq:dyn_all}. Both the dynamics and the control scheme have already been analyzed in \cite{8723153}. In addition to convergence of $x_p$ and $u_s$ to the equilibrium, we want to demonstrate that the bucket fills up by means of the modified terminal cost \eqref{eq:term_cost_total}. In order to make a statement regarding the convergence behavior of the NCS using terminal cost \eqref{eq:term_cost_total}, we rely on \cite[Theorem 1]{DBLP:journals/corr/abs-1902-08132}.

Consider the control sequence of length $q$
\begin{equation} \
\nu_0 =\begin{bmatrix}
u_c^T & 1
\end{bmatrix}^T \quad \text{and} \quad \nu_j=\begin{bmatrix}
0 & 0\end{bmatrix}^T, \quad \forall j \in \mathbb{I}_{[1,q-1]}. 
\label{eq:control_seq}
\end{equation}
Then, the map from an initial plant state $x_{p}$ to the state after $i \in \mathbb{I}_{[0,q]}$ time steps under application of control sequence \eqref{eq:control_seq} to plant \eqref{eq:plant} is recursively defined by
\begin{equation}
f_{p,i}(x_p,u_c)\coloneqq f_p(f_{p,i-1}(x_p,u_c),u_c), \; f_{p,0}(x_p,u_c)\coloneqq x_p.
\label{eq:map_xp}
\nonumber
\end{equation}
With this map in mind, we state the first assumption, which requires the existence of a $q$-step control invariant set for the plant controlled by an input which is held over $q$ time steps (see \cite{8723153}).
\begin{ass}
\label{ass:x_f}
[\cite{8723153}] There exists a closed set $\mathbb{X}_{f,p} \subseteq \mathbb{X}_p$ containing the origin and a $k_p: \mathbb{X}_{f,p} \rightarrow \mathbb{U}_p$ such that for all $x_p \in \mathbb{X}_{f,p}$, it holds that $f_{p,i}(x_p,k_p(x_p)) \in \mathbb{X}_p$ for all $i \in \mathbb{I}_{[1,q-1]}$ and $f_{p,q}(x_p,k_p(x_p)) \in \mathbb{X}_{f,p}$.
\end{ass}
\begin{ass}
\label{ass:v_f}	
[\cite{8723153}] There exists a continuous, positive definite function $V_{f,p}:\mathbb{X}_{f,p}\rightarrow \mathbb{R}$ such that for all $x_p \in \mathbb{X}_{f,p}$, with $V_{f,p},k_p$, $q$ from Assumption \ref{ass:x_f}
\begin{equation*}
\begin{split}
V_{f,p}(f_{p,q}(x_p,&k_p(x_p)))-V_{f,p}(x_p)  \\
\leq & - q \Vert k_p(x_p) \Vert^2_R  -\sum \limits_{i=0}^{q-1}\Vert f_{p,i}(x_p,k_p(x_p)) \Vert^2_Q.
\end{split}
\label{ass:term_reg}
\end{equation*}
\end{ass}
This ensures a decrease of $V_{f,p}$ over a total of $q$ steps under the control sequence \eqref{eq:control_seq}. It allows a temporary increase of the terminal cost as long as there is an overall decrease after $q$ steps.

On the basis of Assumptions \ref{ass:x_f} and \ref{ass:v_f}, we choose the terminal region for the overall state as in \cite{8723153}, i.e.,
\begin{equation}
\begin{split}
\mathbb{X}_f = \underbrace{ \lbrace 0 \rbrace \times \lbrace 0 \rbrace\times \mathbb{I}_{[0,c-g-1]}}_{=: \mathbb{X}_{f}^{\prime}} \cup \underbrace{\mathbb{X}_{f,p} \times \mathbb{U}_p \times \mathbb{I}_{[c-g,b]}}_{=: \mathbb{X}_{f}^{\prime\prime}}. 
\end{split}
\label{eq:term_region}
\end{equation}
To ensure recursive feasibility and convergence, the terminal region $\mathbb{X}_f$ also needs to fulfill $q$-step control invariance. To this end, consider the terminal control sequence
\begin{equation}
\begin{split}
&\begin{aligned}
\kappa_0(x)\coloneqq
\begin{cases}
\begin{bmatrix}
0 & 0\\
\end{bmatrix}^T, & x \in \mathbb{X}_{f}^{\prime}\\
\begin{bmatrix} 
k_p(x_p) &  1
\end{bmatrix}^T, & x \in \mathbb{X}_{f}^{\prime\prime}
\end{cases} 
\end{aligned}\\
&\begin{aligned}
\kappa_j(x)\coloneqq \begin{bmatrix}
0 & 0\\
\end{bmatrix}^T, \quad \forall j \in \mathbb{I}_{[1,q-1]}.
\end{aligned}
\end{split}
\label{eq:term_controller}
\end{equation}
Then we define the map from the initial overall state $x$ to the state after $i$ time steps under application of the terminal control sequence \eqref{eq:term_controller} to system \eqref{eq:dyn_all} by
\begin{equation}
\begin{split}
f_i(x)\coloneqq f(f_{i-1}(x),\kappa_{(i-1)\text{mod}q}(f_{i-1}(x)), \; f_0(x)\coloneqq x.
\label{eq:overall_map}
\end{split}
\end{equation}
We focus now on the terminal cost for the bucket level. To establish convergence in the following, we require a special relation between $c$ and $g$.
\begin{ass}
\label{ass:c_g}
The parameters of the token bucket TS fulfill  $\frac{c}{g} \notin \mathbb{I}$.
\end{ass}
\begin{lemma}
\label{lemma:qg-c}
If Assumption \ref{ass:c_g} holds, then $qg-c\geq 1$.
\end{lemma}
\begin{pf}
Recall that $q = \lceil\frac{c}{g}\rceil$. If $\frac{c}{g} \notin \mathbb{I}$, then $q$ must be strictly larger than $\frac{c}{g}$, i.e., $qg-c > 0$. Since $qg \in \mathbb{I}$ and $c \in \mathbb{I}$, the difference must be an integer larger than zero, or equivalently, $qg-c\geq 1$. \QEDB
\end{pf}
In \eqref{eq:overall_map}, a multi-step map of the overall state $x$ under the terminal control sequence \eqref{eq:term_controller} is defined. We denote by $f_{\beta,i}(x)$ the last row of $f_i(x)$, which pertains to the bucket level under \eqref{eq:term_controller}. Lemma \ref{lemma:alpha} shows that under application of the terminal control sequence \eqref{eq:term_controller}, more tokens are generated than being consumed.
\begin{lemma}
\label{lemma:alpha}
Suppose that Assumption \ref{ass:c_g} holds. Then for all $x\in\mathbb{X}_f$, $V_{f,\beta}(f_{\beta,q}(x)) - V_{f,\beta}(\beta)\leq -\alpha(\beta)$, with $\alpha : \mathbb{I}_{[0,b]} \rightarrow \mathbb{R}$, where $ \alpha(b)=0$ and $\alpha(\beta) > 0$ for all $\beta \in  \mathbb{I}_{[0,b-1]}$.
\end{lemma}
\begin{pf} 
We define
\begin{equation}
z(x)\coloneqq f_{\beta,q}(x)-\beta.
\label{eq:def_z}
\end{equation}
To quantify $z(x)$, we distinguish the situations in which the bucket can hold the full amount of incoming tokens from those where the bucket level is exceeded under the terminal control sequence \eqref{eq:term_controller}. Three different cases are conceivable. In the first case, no transmission 
occurs since $x \in \mathbb{X}_f^\prime$ and the bucket can store the resulting $qg$ arriving tokens, i.e., $z(x)=qg$.
This is true for the following cases:
\begin{itemize}
	\item{$x \in \lbrace 0 \rbrace \times \lbrace 0 \rbrace \times \mathbb{I}_{[0,b-qg-1]}\coloneqq \mathbb{G}_1$ $\subseteq \mathbb{X}_{f}^{\prime}$, if $b-qg\leq c-g-1$ and}
	\item{$x \in \lbrace 0 \rbrace \times \lbrace 0 \rbrace \times \mathbb{I}_{[0,c-g-1]}\coloneqq \mathbb{L} = \mathbb{X}_{f}^{\prime}$, if $b-qg\geq c-g$. } 
\end{itemize}
In the second case, a transmission is triggered and the bucket can hold further $qg-c$ tokens, i.e., $z(x)=qg-c$. This is true for
\begin{itemize}
	\item{$x \in \mathbb{X}_{f,p} \times \mathbb{U}_p \times \mathbb{I}_{[c-g,b-qg+c-1]} \coloneqq \mathbb{H}_1 \subseteq \mathbb{X}_{f}^{\prime\prime}$, if $b-qg+c\geq c-g$ and}
	\item{$x \in \mathbb{X}_{f,p} \times \mathbb{U}_p \times \mathbb{I}_{[c-g,b-qg+c-1]} \coloneqq \mathbb{K}_1 \subseteq \mathbb{X}_{f}^{\prime\prime}$, if $b-qg\geq c-g$.}	
\end{itemize}
In the third case, the bucket does not have enough capacity to store all the arriving tokens. Only the $b-\beta$ tokens missing for full capacity are therefore saved, i.e., $z(x)=b-\beta$. This is valid for
\begin{itemize}
	\item{$x \in \lbrace 0 \rbrace \times \lbrace 0 \rbrace \times \mathbb{I}_{[b-qg,c-g-1]}\coloneqq \mathbb{G}_2$ $\subseteq \mathbb{X}_{f}^{\prime}$, if $b-qg\leq c-g-1$,}	
	\item{$x \in \mathbb{X}_{f,p} \times \mathbb{U}_p \times \mathbb{I}_{[c-g,b]}\coloneqq \mathbb{J} = \mathbb{X}_{f}^{\prime\prime}$, if $b-qg+c\leq c-g-1$,}
	\item{$x \in \mathbb{X}_{f,p} \times \mathbb{U}_p \times \mathbb{I}_{[b-qg+c,b]} \coloneqq \mathbb{H}_2 \subseteq \mathbb{X}_{f}^{\prime\prime}$, if $b-qg+c\geq c-g$ and}
	\item{$x \in \mathbb{X}_{f,p} \times \mathbb{U}_p \times \mathbb{I}_{[b-qg+c,b]} \coloneqq \mathbb{K}_2 \subseteq \mathbb{X}_{f}^{\prime\prime}$, if $b-qg\geq c-g$.}
\end{itemize}
Note that in all three parameter constellations, the entire terminal set is covered. By the above analysis, in the first subcase $b-qg \leq c-g-1$ and $b-qg+c\geq c-g$, $\mathbb{X}_f=\mathbb{G}_1 \cup \mathbb{G}_2 \cup \mathbb{H}_1 \cup \mathbb{H}_2$. In the second subcase $b-qg \leq c-g-1$ and $b-qg+c \leq c-g-1$, it is $\mathbb{X}_f=\mathbb{G}_1 \cup \mathbb{G}_2 \cup \mathbb{J}$. In the last subcase $b-qg\geq c-g$, we have $\mathbb{X}_f=\mathbb{L} \cup \mathbb{K}_1 \cup \mathbb{K}_2$.\\
 We see from the previous calculations that we can define a lower bound for $z(x)$, namely $z(x)\hspace{-1pt}=\hspace{-1pt}0$ if $\beta\hspace{-1pt}=\hspace{-1pt}b$ and $z(x)\hspace{-1pt}\geq \hspace{-1pt}1$ if $\beta \hspace{-1pt}\leq\hspace{-1pt} b-1$. This lower bound holds since $qg > qg-c \geq 1$, due to Assumption \ref{ass:c_g}. By introducing $\zeta(\beta), \zeta: \mathbb{I}_{[0,b]} \rightarrow \mathbb{I}$, $z(x)$ can be described by $z(x) \geq b - \zeta(\beta)$ with
\begin{equation}
 \zeta(\beta) = \begin{cases}
b, & \beta = b \\
b-1, & \beta \in\mathbb{I}_{[0,b-1]}
\end{cases}.
\label{eq:est_z}
\end{equation}
With \eqref{eq:def_z} and \eqref{eq:est_z}, $f_{\beta,q}(x)\geq \beta + b - \zeta(\beta)$ holds. It follows $V_{f,\beta}(f_{\beta,q}(x))-V_{f,\beta}(\beta)= -\sigma(f_{\beta,q}(x)^2-\beta^2)\leq -\sigma((\beta+b-\zeta(\beta))^2-\beta^2)$. If $\beta=b$, $V_{f,\beta}(f_{\beta,q}(x))-V_{f,\beta}(\beta)=0$ holds. If $\beta \in\mathbb{I}_{[0,b-1]}$, this results in $V_{f,\beta}(f_{\beta,q}(x))-V_{f,\beta}(\beta)\leq -\sigma(2\beta+1)\eqqcolon-\alpha(\beta)< 0$.
\QEDB
\end{pf}

\subsection{Convergence}
In the previous section, we introduced all the necessary aspects for the application of \cite[Theorem 1]{DBLP:journals/corr/abs-1902-08132}. In addition, we want to prove that $\beta(k)$ converges to the upper sector of the bucket. These statements are summarized in the following theorem.
\begin{thm}
\label{thm:upper_sector}
Suppose that Assumptions \ref{ass:x_f}-\ref{ass:c_g} hold. Then, if $\mathbb{P}(x(0))$ is feasible,  $\mathbb{P}(x(jM))$ is feasible for all $j \in \mathbb{I}_{\geq 0}$ and $x_p(k)$ and $u_s(k)$ converge to $0$ as $k \rightarrow \infty$. Additionally, $\beta(k)$ converges to the set $\left[\max\lbrace 0,b-Ng\rbrace,b\right]$ as $k \rightarrow \infty$, and the subsequence $\beta(jM)$ converges to $\left[\max\lbrace 0,b-(N-M)g\rbrace,b\right]$ as $j \rightarrow \infty$.		
\end{thm}
   	\vspace*{-2mm}
\begin{pf}
First, we prove convergence of the overall state to the set $\bar{\mathbb{X}}\coloneqq \lbrace 0 \rbrace \times \lbrace 0 \rbrace \times \mathbb{I}_{[0,b]}$ by verifying that \cite[Assumptions 1-4]{DBLP:journals/corr/abs-1902-08132} are fulfilled.\\
The lowest asymptotic average cost $\ell^*_{av}=0$ is attained in the set $\bar{\mathbb{X}}$. With the storage function $\lambda(x)=\Vert u_s \Vert_S^2$ and $R\geq S>0$, the inequality $\ell(x,u)+\lambda(x)-\lambda(f(x,u))-\ell^*_{av} \geq \Vert x_p \Vert_Q^2 + \Vert u_s \Vert_S^2$ holds. Hence, \cite[ Assumption 1]{DBLP:journals/corr/abs-1902-08132} is fulfilled.

The terminal region \eqref{eq:term_region} fulfills \cite[Assumption 2]{DBLP:journals/corr/abs-1902-08132}. For a proof, we refer the reader to the proof of \cite[Theorem 1]{8723153}.

Due to Lemma \ref{lemma:alpha},  $V_{f,\beta}(f_{\beta,q}(x))-V_{f,\beta}(\beta)\leq 0$ for all $x\in\mathbb{X}_f$. With Assumption \ref{ass:v_f}, for all $x \in \mathbb{X}_{f}^{\prime\prime}$ it holds that
\begin{equation}
\begin{split}
&V_f(f_q(x))-V_f(x)
=V_{f,p}(f_{p,q}(x_p,k_p(x_p)))\\&+V_{f,\beta}(f_{\beta,q}(x))-V_{f,p}(x_p)-V_{f,\beta}(\beta)\\ 
&\leq - q \Vert k_p(x_p) \Vert^2_R - \sum \limits_{i=0}^{q-1}\Vert f_{p,i}(x_p,k_p(x_p)) \Vert^2_Q  \\
&= -\sum\limits_{i=0}^{q-1}\ell(f_i(x), \kappa_i(f_i(x))).
\end{split}
\label{eq:check_ass3}
\end{equation}
For all $x \in \mathbb{X}_{f}^{\prime}$, it holds due to $f_p(0,0)=0$ that
\begin{equation}
\begin{split}
&V_f(f_q(x))-V_f(x)\\&=V_{f,p}(0)+V_{f,\beta}(f_{\beta,q}(x))-V_{f,p}(0)-V_{f,\beta}(\beta)\\ &\leq 0 = -\sum\limits_{i=0}^{q-1}\ell(f_i(x), \kappa_i(f_i(x))).
\end{split}
\nonumber
\end{equation}
Since $f_{jq}(x) \in \mathbb{X}_f$, $j \in \left[ 1 ,r \right]$ for all $x\in\mathbb{X}_f$ and $M=rq$, we can repeat these arguments to obtain
\begin{equation*}
V_f(f_M(x))-V_f(x) \le - \sum\limits_{i=0}^{M-1}\ell(f_i(x), \kappa_i(f_i(x))), \quad \forall x\in\mathbb{X}_f,
\end{equation*}
i.e., \cite[ Assumption 3]{DBLP:journals/corr/abs-1902-08132} is fulfilled.

The rotated terminal cost is $\bar{V}_f(x)=V_{f,p}(x_p) + V_{f,\beta}(\beta)+\lambda(x)=V_{f,p}(x_p)+\sigma (b^2-\beta^2)+\Vert u_s \Vert_S^2$. Because $V_{f,p}$ is positive definite, its minimum $\bar{V}_f(x)=0$ is attained on $\lbrace 0 \rbrace \times \lbrace 0 \rbrace \times \lbrace b \rbrace \subset \bar{\mathbb{X}}$. Thus, \cite[Assumption 4]{DBLP:journals/corr/abs-1902-08132} is fulfilled as well.

In summary, all the conditions of \cite[Theorem 1]{DBLP:journals/corr/abs-1902-08132} are met. As a result, the optimization problem $\mathbb{P}$ is feasible for all $k \in \mathbb{I}_{\geq 0}$ and the closed loop state $x(k)$ converges to $\bar{\mathbb{X}}=\lbrace 0 \rbrace \times \lbrace 0 \rbrace \times \mathbb{I}_{[0,b]}$ as $k\rightarrow \infty$. The first part of the result is established.

To obtain the more precise convergence statement for $\beta(k)$, we apply Lemma \ref{lemma:alpha} to obtain a tighter upper bound for the decrease of the terminal cost. We recall that $M=rq$ and $f_{jq}(x) \in \mathbb{X}_f$, $j \in \left[ 1 ,r \right]$, such that it follows for all $x \in \mathbb{X}_f$
\begin{align}
&\begin{aligned}
&V_f(f_M(x))-V_f(x)\le -\sum\limits_{i=0}^{M-1}\ell(f_i(x), \kappa_i(f_i(x)))  \\
&+ V_{f,\beta}(f_{\beta,M}(x))-V_{f,\beta}(f_{\beta,M-q}(x))+ \ldots + V_{f,\beta}(f_{\beta,q}(x))
\nonumber
\end{aligned}\\
&\begin{aligned}
- V_{f,\beta}(\beta)  \stackrel{\text{Lemma 2}}{\leq} -\sum\limits_{i=0}^{M-1}\ell(f_i(x), \kappa_i(f_i(x)))
\label{eq:mod_ass3}
\end{aligned}\\
&\begin{aligned}
&-\alpha(f_{\beta,M-q}(x))-\alpha(f_{\beta,M-2q}(x))-\ldots-\alpha(\beta)\\
&\leq -\sum\limits_{i=0}^{M-1}\ell(f_i(x), \kappa_i(f_i(x)))-\alpha(\beta).
\nonumber
\end{aligned}
\end{align} \\          
We use the rotated optimal control problem $\bar{\mathbb{P}}(x(jM))$ known from economic MPC defined by
\begin{equation*}
\begin{split}
\bar{V}^*(x(jM))\coloneqq \min\limits_{u(\cdot \vert jM)}  \sum\limits_{i=0}^{N-1}&L(x(i \vert jM),u(i \vert jM))\\&+\hspace{-1pt}\bar{V}_f(x(N\vert jM))
\end{split}
\end{equation*}
subject to the same constraints as $\mathbb{P}(x(jM))$, to analyze convergence of $\beta$. Hereby, $L(x,u)\coloneqq \ell(x,u)+\lambda(x)-\lambda(f(x,u))-\ell^*_{av}$ denotes the rotated stage cost and $\bar{V}_f(x)\coloneqq V_f(x)+\lambda(x)$ the rotated terminal cost.\\ 
Assuming that $\bar{\mathbb{P}}$ was feasible at time $jM$, we consider the feasible control input $\tilde{u}(\cdot \vert (j+1)M)$ as
\begin{align}
&\begin{aligned}
\tilde{u}(i \vert (j+1)M))
\nonumber
\end{aligned}\\
&\begin{aligned}
\coloneqq\begin{cases}
u^*(i \vert jM), &\parbox[t]{.6\textwidth}{$i \in [M,N-1]$}\\
\kappa_{(i-N)\text{mod}q}(f_{i-N}(x^*(N\vert jM)))
,  &\parbox[t]{.6\textwidth}{$i \in [N,M+N-1]$.} 
\end{cases}
\nonumber
\end{aligned}
\end{align}
From this we obtain, using standard arguments
\begin{align*}
&\begin{aligned}
&\bar{V}^*(x((j+1)M))-\bar{V}^*(x(jM))\\
\end{aligned}\\
&\begin{aligned}
&\leq - \sum \limits_{i=0}^{M-1}L(x^*(i \vert j M),u^*(i \vert j M)) \\
&+ \sum \limits_{i=0}^{M-1}L(f_i(x^*(N\vert jM)),\kappa_{i}(f_{i}(x^*(N\vert jM))))\\
&+\bar{V}_{f}(f_M(x^*(N\vert jM)))-\bar{V}_{f}(x^*(N \vert j M))\\
\end{aligned}\\
%
%
%
%
&\begin{aligned}
&\leq-\sum \limits_{i=0}^{M-1}L(x^*(i \vert j M),u^*(i \vert j M)) - \alpha(\beta^*(N \vert j M)).
\end{aligned}
\end{align*}
The last inequality follows from \eqref{eq:mod_ass3} and \cite[Lemma 1]{DBLP:journals/corr/abs-1902-08132}. We obtain
\begin{equation}
\begin{split}
\bar{V}^*&(x((j+1)M)) \leq \bar{V}^*(x(jM))\\&\underbrace{-\sum \limits_{i=0}^{M-1}L(x^*(i \vert j M),u^*(i \vert j M)) }_{\leq - L(x(jM),u^*(0 \vert jM))}-\alpha(\beta^*(N \vert j M)).
\end{split}
\nonumber
\end{equation}
Since $L$ is positive definite and $\alpha(b)=0$, $\alpha(\beta) > 0$, for all $\beta \in  \mathbb{I}_{[0,b-1]}$, $\bar{V}^*(x(jM))$ decreases unless $x(jM)\in \bar{\mathbb{X}}$ and $\beta^*(N\vert jM)=b$. Because $\bar{V}_f$ is lower bounded, $\bar{V}^*$ is also lower bounded and therefore, $\bar{V}^*$ must converge to a constant value. Hence, $L\rightarrow 0$ and $\alpha\rightarrow 0$ as $j \rightarrow \infty$. Due to the positive definiteness of $L$ and $\alpha(b)=0$, $\alpha(\beta) > 0$, for all $\beta \in  \mathbb{I}_{[0,b-1]}$, $x(jM)\rightarrow \bar{\mathbb{X}}$ and $\beta^*(N\vert jM)\rightarrow b$ as $j \rightarrow \infty$. The first $M$ parts of the optimal predicted input are applied in closed loop. From the convergence of $\beta^*(N\vert jM)$ to $b$, it therefore follows that in closed loop, $\beta(k)$ converges to $\left[\max\{0,b-Ng\},b\right]$ as 
$k \rightarrow \infty$ due to the bucket dynamics \eqref{eq:beta_dyn}. Further, the subsequence $\beta(jM)$ converges to $\left[\max\{0,b-(N-M)g\},b\right]$ as $j \rightarrow \infty$. \QEDB	
\end{pf}

\begin{remark}
From Theorem \ref{thm:upper_sector}, the exact convergence sector of $\beta$ depends on the prediction horizon $N$ and the bucket parameters $b$ and $g$, i.e., it depends on these parameters whether the obtained statement is weak or strong. For $N=M$, one can for instance guarantee that $\beta(jM)\rightarrow b$ as $j\rightarrow\infty$, i.e., the bucket level at sampling instances goes exactly to the upper rim.  
\end{remark}
\begin{remark}
From the proof of Theorem \ref{thm:upper_sector}, we can see that with standard MPC tools, only convergence of the plant state and saved input can be established. By the means of the non-standard analysis in this proof, it is possible to make a statement in this respect also for the bucket level.
\end{remark}

\section{Additional Direct Link in the Network }
\label{sec:periodic}
In the previous section, we have shown that, if $\frac{c}{g}$ is not an integer, the bucket level in closed loop lies a above a certain threshold as $k \rightarrow \infty$. This statement may not be very strong under certain parameter constellations (high $N$ and $g$, low $b$), and for some, no statement can be made at all ($\frac{c}{g} \in \mathbb{I}$). To obtain stronger results in this respect, we consider a slightly modified version of the NCS in this section. For this extended setup, we prove convergence of the overall state $x$ to $ \left[ 0,0,b\right]^T $. On top of convergence, we prove asymptotic stability.

\subsection{Modified Setup}
\begin{figure}[!h]
	\centering
	\resizebox {0.5\textwidth} {!} {
		\tikzset{every picture/.style={line width=0.75pt}} 

\begin{tikzpicture}[x=0.75pt,y=0.75pt,yscale=-0.8,xscale=1]

\draw   (160,18) .. controls (160,13.58) and (163.58,10) .. (168,10) -- (232,10) .. controls (236.42,10) and (240,13.58) .. (240,18) -- (240,42) .. controls (240,46.42) and (236.42,50) .. (232,50) -- (168,50) .. controls (163.58,50) and (160,46.42) .. (160,42) -- cycle ;
\draw   (310,18) .. controls (310,13.58) and (313.58,10) .. (318,10) -- (352,10) .. controls (356.42,10) and (360,13.58) .. (360,18) -- (360,42) .. controls (360,46.42) and (356.42,50) .. (352,50) -- (318,50) .. controls (313.58,50) and (310,46.42) .. (310,42) -- cycle ;
\draw   (430,18) .. controls (430,13.58) and (433.58,10) .. (438,10) -- (512,10) .. controls (516.42,10) and (520,13.58) .. (520,18) -- (520,42) .. controls (520,46.42) and (516.42,50) .. (512,50) -- (438,50) .. controls (433.58,50) and (430,46.42) .. (430,42) -- cycle ;
\draw   (250,136) .. controls (250,131.58) and (253.58,128) .. (258,128) -- (432,128) .. controls (436.42,128) and (440,131.58) .. (440,136) -- (440,160) .. controls (440,164.42) and (436.42,168) .. (432,168) -- (258,168) .. controls (253.58,168) and (250,164.42) .. (250,160) -- cycle ;
\draw    (158,30.04) -- (109.5,31) ;

\draw [shift={(160,30)}, rotate = 178.87] [fill={rgb, 255:red, 0; green, 0; blue, 0 }  ][line width=0.75]  [draw opacity=0] (8.93,-4.29) -- (0,0) -- (8.93,4.29) -- cycle    ;
\draw    (110,150) -- (109.5,31) ;

\draw    (110,150) -- (250,150) ;

\draw    (308,30) -- (240,30) ;

\draw [shift={(310,30)}, rotate = 180] [fill={rgb, 255:red, 0; green, 0; blue, 0 }  ][line width=0.75]  [draw opacity=0] (8.93,-4.29) -- (0,0) -- (8.93,4.29) -- cycle    ;
\draw    (428,30) -- (360,30) ;

\draw [shift={(430,30)}, rotate = 180] [fill={rgb, 255:red, 0; green, 0; blue, 0 }  ][line width=0.75]  [draw opacity=0] (8.93,-4.29) -- (0,0) -- (8.93,4.29) -- cycle    ;
\draw    (440,150) -- (560,150) ;

\draw    (520,30) -- (560,30) ;

\draw    (560,30) -- (560,94) ;

\draw    (560,120) -- (566.5,127) ;

\draw    (560,130) -- (560,150) ;

\draw    (560,100) -- (560,120) ;

\draw    (560,94) -- (554.5,100) ;

\draw    (112,100) -- (554.5,100) ;

\draw [shift={(110,100)}, rotate = 0] [fill={rgb, 255:red, 0; green, 0; blue, 0 }  ][line width=0.75]  [draw opacity=0] (8.93,-4.29) -- (0,0) -- (8.93,4.29) -- cycle    ;
\draw  [dash pattern={on 0.84pt off 2.51pt}] (80,70) -- (500,70) -- (500,180) -- (80,180) -- cycle ;

\draw (200,30) node  [align=left] {Actuator};
\draw (335,30) node  [align=left] {Plant};
\draw (475,30) node  [align=left] {Controller};
\draw (345,148) node  [align=left] {Token Bucket TS};
\draw (274,21) node  [align=left] {$u_p$};
\draw (389.5,21) node  [align=left] {$x_p$};
\draw (540.5,21) node  [align=left] {$u_c$};
\draw (545,129) node  [align=left] {$\gamma(k)$};
\draw (575,100) node  [align=left] {$\delta(k)$};
\draw (110.5,169) node  [align=left] {Network};
\draw (345,90) node  [align=left] {Direct Link};

\end{tikzpicture}
	}
	\caption[{Modified NCS configuration}]{Modified NCS with additional direct link.}
	\label{fig:NCS_config_periodic}
\end{figure}
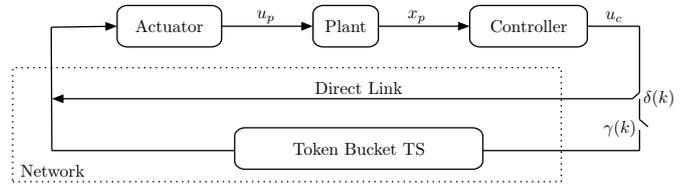

We modify the already introduced NCS setup by adding a direct link between controller and actuator as seen in Figure \ref{fig:NCS_config_periodic}. This direct link enables transmissions which do not need to fulfill the token bucket TS and are, for instance, periodically triggered.
Whether a transmission passes the direct link or needs to fulfill the token bucket TS is determined by the binary transmission sequence $\delta(\cdot)$. Here, $\delta(k)=0$ indicates a transmission which must fulfill the constraints given by the token bucket whereas if $\delta(k)=1$, the control input $u_c$ is transmitted over the direct link. In general, $\delta(k)+\gamma(k)\leq 1$ for all $k \in \mathbb{I}_{\geq 0}$, since it is not possible to have multiple transmissions through the network at the same time. In the modified setup, the applied input becomes
\begin{equation}
u_p(k)=(\gamma(k)+\delta(k))u_c(k)+(1-\gamma(k)-\delta(k))u_p(k-1)
\nonumber
\end{equation} 
with $u_p(-1) \in \mathbb{U}_p$. As transmissions via the direct link do not need to fulfill the token bucket TS, they do not consume any tokens. Moreover, at times when $\delta(k)=1$, no tokens are added to the bucket. The right-hand side of the overall dynamics (\ref{eq:dyn_all}) with the extended overall control $u\coloneqq \left[ u_c^T,\gamma, \delta \right]^T$ is then characterized by
\begin{equation}
f(x,u)\coloneqq\begin{bmatrix}
f_p(x_p,(\gamma+\delta) u_c+(1-\gamma-\delta)u_s)\\(\gamma+\delta) u_c+(1-\gamma-\delta)u_s\\\min\lbrace \beta + (1-\delta) g - \gamma c,b \rbrace
\end{bmatrix}
\nonumber
\label{eq:dyn_matrix_periodic}
\end{equation}
with state $x(k)\coloneqq \mathbb{X}_p \times \mathbb{U}_p \times \mathbb{I}_{[0,b]} \eqqcolon\mathbb{X}\subseteq \mathbb{R}^{n_p+m_p+1}$ and input $u(k)\coloneqq \mathbb{U}_p \times \Gamma\Delta \eqqcolon\mathbb{U} \subseteq \mathbb{R}^{m_p+2}$, $\Gamma\Delta\coloneqq \lbrace \gamma,\delta \in \lbrace0,1\rbrace \vert \gamma+\delta\leq 1 \rbrace$.

The stage cost associated with the NCS reads
\begin{equation}
\ell(x,u)=\Vert x_p \Vert^2_Q + (\gamma+\delta)\Vert u_c\Vert^2_R +(1-\gamma-\delta) \Vert u_s \Vert^2_R+\ell_\beta(\beta)
\nonumber
\label{eq:stage_cost_periodic}
\end{equation}
with $\ell_\beta(\beta)\coloneqq\psi(b^2-\beta^2)$, $\psi \in (0,\infty)$. In addition to the performance index of the plant, now there is the positive definite term $\ell_\beta$ penalizing the deviation of the current filling level from the maximum capacity of the bucket. This is necessary to guarantee exact convergence of $\beta$ to the upper rim of the bucket. The appropriate choice of the scaling parameter $\psi$ will be discussed later. For the terminal cost, we apply the same cost as in the previous section, namely $V_f(x) = V_{f,p}(x_p) + V_{f,\beta}(\beta)$ with $V_{f,\beta}(\beta)=\sigma (b^2-\beta^2)$.
\begin{remark}
In the setup of Section \ref{sec:extended_term}, the proof of convergence as in Theorem \ref{thm:upper_sector} fails if we added the term $\ell_\beta$ already there. Notice that therein, $V_{f,\beta}(f_{\beta,q}(x))-V_{f,\beta}(\beta)\leq 0$ was sufficient to prove convergence under Assumption \ref{ass:v_f}, since no stage cost on the bucket was considered. However, we have $V_{f,\beta}(f_{\beta,q}(\left[x_p^T \: u_s^T \: b \right]^T ))-V_{f,\beta}(b)=0$ and $\sum_{i=0}^{q-1}\ell_\beta(f_{\beta,i}(\left[x_p^T \: u_s^T \: b \right]^T ))> 0$ under the terminal control sequence (\ref{eq:term_controller}), such that $V_{f,\beta}(f_{\beta,q}(x))-V_{f,\beta}(\beta) \leq - \sum_{i=0}^{q-1}\ell_\beta(f_{\beta,i}(x))$ is not true in the entire terminal region. Hence, Assumption \ref{ass:v_f} would not be sufficient for convergence anymore.
\end{remark}

\subsection{Periodic Transmission over the Direct Link}
\begin{figure}[!h]
	\centering
	\resizebox {0.5\textwidth} {!} {
		\tikzset{every picture/.style={line width=0.75pt}} 

\begin{tikzpicture}[x=0.75pt,y=0.75pt,yscale=-1,xscale=1]

\draw    (120,120) -- (210,120) ;
\draw [shift={(210,120)}, rotate = 0] [color={rgb, 255:red, 0; green, 0; blue, 0 }  ][fill={rgb, 255:red, 0; green, 0; blue, 0 }  ][line width=0.75]      (0, 0) circle [x radius= 3.35, y radius= 3.35]   ;
\draw [shift={(120,120)}, rotate = 0] [color={rgb, 255:red, 0; green, 0; blue, 0 }  ][fill={rgb, 255:red, 0; green, 0; blue, 0 }  ][line width=0.75]      (0, 0) circle [x radius= 3.35, y radius= 3.35]   ;
\draw    (210,120) -- (300,120) ;
\draw [shift={(300,120)}, rotate = 0] [color={rgb, 255:red, 0; green, 0; blue, 0 }  ][fill={rgb, 255:red, 0; green, 0; blue, 0 }  ][line width=0.75]      (0, 0) circle [x radius= 3.35, y radius= 3.35]   ;
\draw [shift={(210,120)}, rotate = 0] [color={rgb, 255:red, 0; green, 0; blue, 0 }  ][fill={rgb, 255:red, 0; green, 0; blue, 0 }  ][line width=0.75]      (0, 0) circle [x radius= 3.35, y radius= 3.35]   ;
\draw    (300,120) -- (390,120) ;
\draw [shift={(390,120)}, rotate = 0] [color={rgb, 255:red, 0; green, 0; blue, 0 }  ][fill={rgb, 255:red, 0; green, 0; blue, 0 }  ][line width=0.75]      (0, 0) circle [x radius= 3.35, y radius= 3.35]   ;
\draw [shift={(300,120)}, rotate = 0] [color={rgb, 255:red, 0; green, 0; blue, 0 }  ][fill={rgb, 255:red, 0; green, 0; blue, 0 }  ][line width=0.75]      (0, 0) circle [x radius= 3.35, y radius= 3.35]   ;
\draw    (390,120) -- (438,120) ;
\draw [shift={(440,120)}, rotate = 180] [fill={rgb, 255:red, 0; green, 0; blue, 0 }  ][line width=0.75]  [draw opacity=0] (8.93,-4.29) -- (0,0) -- (8.93,4.29) -- cycle    ;
\draw    (80,120) -- (120,120) ;

\draw    (120,120) -- (150,120) ;
\draw [shift={(150,120)}, rotate = 45] [color={rgb, 255:red, 0; green, 0; blue, 0 }  ][line width=0.75]    (-5.59,0) -- (5.59,0)(0,5.59) -- (0,-5.59)   ;

\draw    (150,120) -- (180,120) ;
\draw [shift={(180,120)}, rotate = 45] [color={rgb, 255:red, 0; green, 0; blue, 0 }  ][line width=0.75]    (-5.59,0) -- (5.59,0)(0,5.59) -- (0,-5.59)   ;
\draw [shift={(150,120)}, rotate = 45] [color={rgb, 255:red, 0; green, 0; blue, 0 }  ][line width=0.75]    (-5.59,0) -- (5.59,0)(0,5.59) -- (0,-5.59)   ;
\draw    (180,120) -- (193.5,120) -- (210,120) ;

\draw [shift={(180,120)}, rotate = 45] [color={rgb, 255:red, 0; green, 0; blue, 0 }  ][line width=0.75]    (-5.59,0) -- (5.59,0)(0,5.59) -- (0,-5.59)   ;
\draw    (210,120) -- (240,120) ;

\draw    (240,120) -- (270,120) ;
\draw [shift={(270,120)}, rotate = 45] [color={rgb, 255:red, 0; green, 0; blue, 0 }  ][line width=0.75]    (-5.59,0) -- (5.59,0)(0,5.59) -- (0,-5.59)   ;

\draw    (270,120) -- (300,120) ;

\draw [shift={(270,120)}, rotate = 45] [color={rgb, 255:red, 0; green, 0; blue, 0 }  ][line width=0.75]    (-5.59,0) -- (5.59,0)(0,5.59) -- (0,-5.59)   ;
\draw    (300,120) -- (330,120) ;
\draw [shift={(330,120)}, rotate = 45] [color={rgb, 255:red, 0; green, 0; blue, 0 }  ][line width=0.75]    (-5.59,0) -- (5.59,0)(0,5.59) -- (0,-5.59)   ;

\draw    (330,120) -- (360,120) ;

\draw [shift={(330,120)}, rotate = 45] [color={rgb, 255:red, 0; green, 0; blue, 0 }  ][line width=0.75]    (-5.59,0) -- (5.59,0)(0,5.59) -- (0,-5.59)   ;
\draw    (360,120) -- (390,120) ;

\draw (119.5,139) node  [align=left] {$jq$};
\draw (219.5,139) node  [align=left] {$(j+1)q$};
\draw (302,139) node  [align=left] {$\cdots$};

\end{tikzpicture}
	}

	\caption{Transmission sequence with periodic transmissions ($\bullet$) and transmissions which fulfill the token bucket TS ($\times$).}
	\label{fig:trans_seq}
\end{figure}
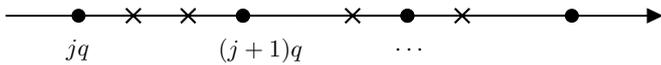

We consider a periodic transmission sequence $\delta(\cdot)$ over the direct link
\begin{equation}
\delta(k)=\begin{cases}
1, \quad \text{if} \quad  k=jq, \quad j \in \mathbb{I}_{\geq 0}\\
0, \quad \text{else}.
\end{cases}
\nonumber
\end{equation}
These transmissions are guaranteed to be possible and may be triggered every $jq$ time steps. Combined with transmissions which fulfill the token bucket TS, a possible transmission sequence is visualized in Figure \ref{fig:trans_seq}.
In the following, we consider this type of mixed transmission specification to prove convergence and asymptotic stability of the overall state $x$. 

\begin{remark}
This transmission pattern is inspired by the time slotted communication abstraction as introduced in \cite{8651811}. Periodic transmissions over the direct link pertain to deterministic transmissions, whereas the remaining transmissions, which need to fulfill the token bucket TS, correspond to opportunistic ones.
\end{remark}

\subsection{Guarantee of Convergence}
By verifying the conditions of \cite[Theorem 1]{DBLP:journals/corr/abs-1902-08132} we prove convergence of the overall state $x$ to  $\left[ 0,0,b\right]^T$. For convenience, we apply again \cite[Theorem 1]{DBLP:journals/corr/abs-1902-08132}, although we only want to prove convergence to a set point. Alternatively, methods as in \cite{amrit2011economic} can be considered here.

We assume that Assumptions 1 and 2 of the previous section hold and define the terminal region $\mathbb{X}_f$ as
\begin{equation}
\mathbb{X}_f\coloneqq \mathbb{X}_{f,p} \times \mathbb{U}_{p}\times \mathbb{I}_{[0,b]}.
\label{eq:term_region_periodic}
\end{equation}
With respect to this terminal region, we consider the terminal control sequence
\begin{equation}
\begin{split}
\kappa_0(x)&\coloneqq\begin{bmatrix}
k_p(x_p) & 0 & 1
\end{bmatrix}^T\\
\text{and}  \quad \kappa_j(x)&\coloneqq\begin{bmatrix}
0&0&0\end{bmatrix}^T\hspace{-1pt},\quad \forall j \in \mathbb{I}_{[1,q-1]}.
\end{split} 
\label{eq:term_controller_periodic}
\end{equation}
To have both periodic transmissions over the direct link and transmissions which fulfill the token bucket TS, $q \geq 2$ holds in the following. Note that $q \geq 2$ was implicitly assumed in Section \ref{sec:extended_term} as well by Assumption \ref{ass:c_g}.

\begin{lemma}
	\label{lemma:no_trans}
Under application of the terminal control sequence \eqref{eq:term_controller_periodic}, for all $x\in\mathbb{X}_f$, for $q \geq 2$,  $V_{f,\beta}(f_{\beta,q}(x)) - V_{f,\beta}(\beta)\leq - \alpha(\beta)$ with $\alpha$ as in Lemma \ref{lemma:alpha} holds.
\end{lemma}
\begin{pf}
Equivalently to Lemma \ref{lemma:alpha}, we define $z(x)\coloneqq f_{\beta,q}(x)-\beta$. Since there is only a transmission over the direct link under application of \eqref{eq:term_controller_periodic}, either  $z(x)=(q-1)g$ or  $z(x)=b-\beta$ holds. We denote again $z(x) \geq b- \zeta(\beta)$ with $\zeta(\beta)$ as in (\ref{eq:est_z}). Therefore, $f_{\beta,q}(x)\geq \beta + b - \zeta(\beta)$ holds and $V_{f,\beta}(f_{\beta,q}(x))-V_{f,\beta}(\beta)= -\sigma(f_{\beta,q}(x)^2-\beta^2)\leq -\sigma((\beta+b-\zeta(\beta))^2-\beta^2)$ follows. With $\beta=b$, $V_{f,\beta}(f_{\beta,q}(x))-V_{f,\beta}(\beta)=0$ holds. If $\beta \leq b-1$, this results in $V_{f,\beta}(f_{\beta,q}(x))-V_{f,\beta}(\beta)\leq -\sigma(2\beta+1)\eqqcolon\alpha(\beta)$.\QEDB
\end{pf}
 \begin{lemma}
 \label{lemma:psi_periodic}
Assume that $x\in\mathbb{X}_{f,p}\times\mathbb{U}_p\times\{0\}$ and $(q-1)g\leq b$. Then, for $q \geq 2$, the summed up stage cost $\ell_\beta$ under application of (\ref{eq:term_controller_periodic}) is 
\begin{equation}
\sum\limits_{i=0}^{q-1}\ell_\beta(f_{\beta,i}(x))=
\psi(qb^2-\frac{1}{6}g^2(q-1)(q-2)(2q-3)).
\nonumber
 \end{equation}
 \end{lemma}
   	\vspace*{-2mm}
 \begin{pf}
Since there is no transmission over the token bucket part of the network under application of (\ref{eq:term_controller_periodic}), $f_{\beta,i}(x)=(i-1)g$ with $i\in \mathbb{I}_{[1,q-1]}$ holds due to $\beta = 0$ and $(q-1)g\leq b$. It follows
\begin{equation}
\begin{split}
\sum\limits_{i=0}^{q-1}\ell_\beta(f_{\beta,i}(x))&=\psi q b^2-\psi g^2 \sum \limits_{i=1}^{q-1}(i-1)^2\\
&=\psi(qb^2-\frac{1}{6}g^2(q-1)(q-2)(2q-3)). \quad \QEDB\\
\end{split}
\nonumber
\end{equation}
 \end{pf}
\begin{ass}
	\label{ass:sigma}
The scaling parameters $\sigma$ and $\psi$ are chosen such that $\sigma \geq \psi\left(qb^2-\frac{1}{6}g^2(q-1)(q-2)(2q-3)\right)$.
\end{ass}
With this choice of scaling parameters, we establish convergence in the following theorem.
\begin{thm}
Suppose that Assumptions \ref{ass:x_f}, \ref{ass:v_f} and \ref{ass:sigma} hold and that $q \geq 2$. Then, if $\mathbb{P}(x(0))$ is feasible,  $\mathbb{P}(x(jM))$ is feasible for all $j \in \mathbb{I}_{\geq 0}$ and the overall state $x$ converges to  $\left[ 0,0,b\right]^T$ as $k \rightarrow \infty$.
\end{thm}
\begin{pf}
To prove Theorem 2, we verify the conditions of \cite[Theorem 1]{DBLP:journals/corr/abs-1902-08132} and establish convergence to the ``set'' $\bar{\mathbb{X}}\coloneqq\lbrace \left[ 0,0,b\right]^T \rbrace$.

It is $\ell^*_{av}=0$ in $\bar{\mathbb{X}}$. Hence, \cite[Assumption 1]{DBLP:journals/corr/abs-1902-08132} is fulfilled with $\lambda(x)=\Vert u_s \Vert_S,R\geq S > 0$, since $\ell(x,u)+\lambda(x)-\lambda(f(x,u))-\ell^*_{av} \geq \Vert x_p \Vert_Q^2 + \Vert u_s \Vert_S^2+\psi(b^2-\beta^2)$ holds.

The terminal region (\ref{eq:term_region_periodic}) fulfills \cite[Assumption 2]{DBLP:journals/corr/abs-1902-08132}. With (\ref{eq:term_controller_periodic}), the plant state $x_p$ remains in the constraint set and returns to $\mathbb{X}_{f,p}$ within $q$ steps due to Assumption \ref{ass:term_reg}. Regardless of bucket filling level, applying (\ref{eq:term_controller_periodic}) is always possible, since we do not transmit over the token bucket part of the network. The saved input $u_s$ remains in $\mathbb{U}_p$.

To check \cite[Assumption 3]{DBLP:journals/corr/abs-1902-08132}, we must, due to Assumption 2, merely prove that for all $x\in\mathbb{X}_f$, $V_{f,\beta}(f_{\beta,q}(x))-V_{f,\beta}(\beta)\leq -\sum_{i=0}^{q-1}\ell_\beta(f_{\beta,i}(x))$.
To do this, we consider the following two cases:
\begin{itemize}
	\item[1.)]{$x \in \mathbb{X}_{f,p} \times \mathbb{U}_p \times \lbrace b \rbrace$: With Lemma \ref{lemma:no_trans}, we have
		\begin{equation*}
		V_{f,\beta}(f_{\beta,q}(x))-V_{f,\beta}(\beta)=0=\sum\limits_{i=0}^{q-1} \psi (b^2-f_{\beta,i}(x)^2)
		\end{equation*}
		since $f_{\beta,i}(x)=b$, $i\in[0,q-1]$.}
	\item[2.)]{$x \in \mathbb{X}_{f,p} \times \mathbb{U}_p \times \mathbb{I}_{[0,b-1]}$: We upper bound the decrease of the terminal cost on the bucket level using Lemma \ref{lemma:no_trans} and Assumption \ref{ass:sigma} by
		\begin{equation}
		\label{eq:th2_vf_periodic}
		\begin{split}
			&V_{f,\beta}(f_{\beta,q}(x))-V_{f,\beta}(\beta)\leq-\sigma(2\beta+1) \leq - \sigma \\		
			&\leq -\psi(qb^2-\frac{1}{6}g^2(q-1)(q-2)(2q-3)).
		\end{split}
		\end{equation}	
		Under the terminal control sequence (\ref{eq:term_controller_periodic}), it is apparent that the summed up stage cost $\ell_\beta$ is highest if the bucket is initially empty, i.e., $\beta=0$ and all further arriving tokens can be stored, i.e., $(q-1)g\leq b$. Therefore, using Lemma \ref{lemma:psi_periodic} results in 
		\begin{equation}
		\label{eq:lower_bound_ass3_periodic}
		\begin{split}
		-\sum\limits_{i=0}^{q-1}&\ell_\beta(f_{\beta,i}(x))\geq -\sum\limits_{i=0}^{q-1}\ell_\beta(f_{\beta,i}(\begin{bmatrix}
		x_p^T & u_s^T & 0
		\end{bmatrix}^T))\\ &\geq -\psi q b^2+\psi g^2 \sum \limits_{i=1}^{q-1}(i-1)^2\\
&=-\psi(qb^2-\frac{1}{6}g^2(q-1)(q-2)(2q-3)).
		\end{split}
		\end{equation}}
\end{itemize}
In summary, \cite[Assumption 3]{DBLP:journals/corr/abs-1902-08132} holds. Since $V_{f,p}(x)$ is positive definite, the minimum of $\bar{V}_f(x)$ is attained on $\bar{\mathbb{X}}=\lbrace \left[ 0,0,b\right]^T \rbrace$. Thus, \cite[Assumption 4]{DBLP:journals/corr/abs-1902-08132} is fulfilled as well.

Hence, all the conditions of \cite[Theorem 1]{DBLP:journals/corr/abs-1902-08132} are fulfilled. The optimization problem $\mathbb{P}(x(jM))$ is feasible for all $j \in \mathbb{I}_{\geq 0}$ and the closed loop state $x(k)$ converges to $\bar{\mathbb{X}}=\lbrace \left[ 0,0,b\right]^T \rbrace$ as $k \rightarrow \infty$.
\QEDB
\end{pf}
\begin{remark}
If Assumptions \ref{ass:x_f} and \ref{ass:v_f} hold and both $\ell_\beta=0$ and $V_{f,\beta}=0$, then the overall state $x$ converges to  $\lbrace 0 \rbrace \times \lbrace 0 \rbrace \times \mathbb{I}_{[0,b]}$ as $k \rightarrow \infty$.	
\end{remark}

\subsection{Asymptotic Stability}
In addition, we obtain asymptotic stability of $\left[ 0,0,b\right]^T$ under the conditions of the following theorem.
\begin{thm}
	Suppose that Assumptions \ref{ass:x_f}, \ref{ass:v_f} and  \ref{ass:sigma} hold, $g\ge 2$, that $N=JM$, $J \in \mathbb{I}_{\geq 1}$, $\mathbb{U}_p$ is compact and $(0,0)\in\text{int}(\mathbb{X}_{f,p} \times \mathbb{U}_p)$. Then, if $\mathbb{P}(x(0))$ is feasible,  $\mathbb{P}(x(jM))$ is feasible for all $j \hspace{-1.5pt}\in \hspace{-1.5pt}\mathbb{I}_{\geq 0}$ and $\left[0,0,b\right]^T $ is asymptotically stable.
	
\end{thm}
\begin{pf}
Since $\mathbb{U}_p$ is compact, $\mathbb{U}$ is compact such that \cite[Assumption 5]{DBLP:journals/corr/abs-1902-08132} holds. Hence, the claim follows directly from \cite[Theorem 3]{DBLP:journals/corr/abs-1902-08132}.	\QEDB
\end{pf}

\section{Numerical Example}
\label{sec:num_example}
We illustrate the theoretical results of this paper on the linearized batch reactor taken from \cite{898792}, which was discretized with a sampling time of $0.1$s. The initial plant state is $x_p(0)=\left[ 1, 0, 1, 0 \right]^T$ and the initial saved input is $u_s(0)=\left[ 0, 0 \right]^T$. Both the plant state and saved input are unconstrained. The cost matrices are chosen to $Q=10I$ and $R=I$. We choose $\mathbb{X}_{f,p}=\mathbb{R}^n$ and $V_{f,p}(x_p)=\Vert x_p \Vert_P^2$, where $P$ is according to \cite[Lemma 2]{8723153}, with which Assumptions 1 and 2 are satisfied. For the token bucket, we set $b=22$, $c=8$ and $g=3$, where $\frac{c}{g}=\frac{8}{3} \notin \mathbb{I}$, such that Assumption 3 is satisfied. The rollout controller operates with $N=M=q=3$, such that $r=1$. At the beginning, the bucket is completely filled with $\beta(0)=22$ tokens.

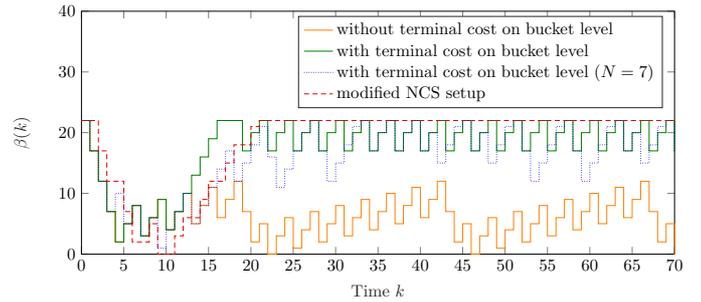
\begin{figure}[!h]
	\centering
	\resizebox {0.5\textwidth} {!} {
%
%
%
\begin{tikzpicture}

\begin{axis}[%
width=4.844in,
height=2in,
at={(0.813in,0.378in)},
scale only axis,
xmin=0,
xmax=70,
xlabel style={font=\color{white!15!black}},
xlabel={Time $k$},
ymin=0,
ymax=40,
ylabel style={font=\color{white!15!black}},
ylabel={$\beta(k)$},
axis background/.style={fill=white},
legend style={legend cell align=left, align=left, draw=white!15!black}
]
\addplot[const plot, color=istorange] table[row sep=crcr] {%
0	22\\
1	17\\
2	12\\
3	7\\
4	2\\
5	5\\
6	8\\
7	3\\
8	6\\
9	9\\
10	4\\
11	7\\
12	10\\
13	5\\
14	8\\
15	11\\
16	6\\
17	9\\
18	12\\
19	7\\
20	2\\
21	5\\
22	0\\
23	3\\
24	6\\
25	1\\
26	4\\
27	7\\
28	2\\
29	5\\
30	8\\
31	3\\
32	6\\
33	9\\
34	4\\
35	7\\
36	10\\
37	5\\
38	8\\
39	11\\
40	6\\
41	9\\
42	12\\
43	7\\
44	2\\
45	5\\
46	0\\
47	3\\
48	6\\
49	1\\
50	4\\
51	7\\
52	2\\
53	5\\
54	8\\
55	3\\
56	6\\
57	9\\
58	4\\
59	7\\
60	10\\
61	5\\
62	8\\
63	11\\
64	6\\
65	9\\
66	12\\
67	7\\
68	2\\
69	5\\
70	0\\
71	3\\
72	6\\
73	1\\
74	4\\
};
\addlegendentry{without terminal cost on bucket level}

\addplot[const plot, color=istgreen] table[row sep=crcr] {%
0	22\\
1	17\\
2	12\\
3	7\\
4	2\\
5	5\\
6	8\\
7	3\\
8	6\\
9	9\\
10	4\\
11	7\\
12	10\\
13	13\\
14	16\\
15	19\\
16	22\\
17	22\\
18	22\\
19	17\\
20	20\\
21	22\\
22	17\\
23	20\\
24	22\\
25	17\\
26	20\\
27	22\\
28	17\\
29	20\\
30	22\\
31	17\\
32	20\\
33	22\\
34	17\\
35	20\\
36	22\\
37	17\\
38	20\\
39	22\\
40	17\\
41	20\\
42	22\\
43	17\\
44	20\\
45	22\\
46	17\\
47	20\\
48	22\\
49	17\\
50	20\\
51	22\\
52	17\\
53	20\\
54	22\\
55	17\\
56	20\\
57	22\\
58	17\\
59	20\\
60	22\\
61	17\\
62	20\\
63	22\\
64	17\\
65	20\\
66	22\\
67	17\\
68	20\\
69	22\\
70	17\\
71	20\\
72	22\\
73	17\\
74	20\\
};
\addlegendentry{with terminal cost on bucket level}

\addplot[const plot, densely dotted ,color=istblue] table[row sep=crcr] {%
0	22\\
1	17\\
2	12\\
3	7\\
4	10\\
5	5\\
6	8\\
7	3\\
8	6\\
9	1\\
10	4\\
11	7\\
12	10\\
13	5\\
14	8\\
15	11\\
16	14\\
17	17\\
18	12\\
19	15\\
20	18\\
21	21\\
22	16\\
23	11\\
24	14\\
25	17\\
26	20\\
27	22\\
28	17\\
29	12\\
30	15\\
31	18\\
32	21\\
33	22\\
34	17\\
35	20\\
36	22\\
37	17\\
38	20\\
39	22\\
40	17\\
41	20\\
42	15\\
43	18\\
44	21\\
45	22\\
46	17\\
47	20\\
48	15\\
49	18\\
50	21\\
51	22\\
52	17\\
53	12\\
54	15\\
55	18\\
56	21\\
57	22\\
58	17\\
59	12\\
60	15\\
61	18\\
62	21\\
63	22\\
64	17\\
65	20\\
66	15\\
67	18\\
68	21\\
69	22\\
70	17\\
71	20\\
72	15\\
73	18\\
74	21\\
};
\addlegendentry{with terminal cost on bucket level $(N=7)$}

\addplot[const plot,densely dashed, color=istred] table[row sep=crcr] {%
	0	22\\
	1	22\\
	2	17\\
	3	12\\
	4	12\\
	5	7\\
	6	2\\
	7	2\\
	8	5\\
	9	0\\
	10	0\\
	11	3\\
	12	6\\
	13	6\\
	14	9\\
	15	12\\
	16	12\\
	17	15\\
	18	18\\
	19	18\\
	20	21\\
	21	22\\
	22	22\\
	23	22\\
	24	22\\
	25	22\\
	26	22\\
	27	22\\
	28	22\\
	29	22\\
	30	22\\
	31	22\\
	32	22\\
	33	22\\
	34	22\\
	35	22\\
	36	22\\
	37	22\\
	38	22\\
	39	22\\
	40	22\\
	41	22\\
	42	22\\
	43	22\\
	44	22\\
	45	22\\
	46	22\\
	47	22\\
	48	22\\
	49	22\\
	50	22\\
	51	22\\
	52	22\\
	53	22\\
	54	22\\
	55	22\\
	56	22\\
	57	22\\
	58	22\\
	59	22\\
	60	22\\
	61	22\\
	62	22\\
	63	22\\
	64	22\\
	65	22\\
	66	22\\
	67	22\\
	68	22\\
	69	22\\
	70	22\\
	71	22\\
	72	22\\
	73	22\\
	74	22\\
	75	22\\
	76	22\\
	77	22\\
	78	22\\
	79	22\\
	80	22\\
	81	22\\
	82	22\\
	83	22\\
	84	22\\
	85	22\\
	86	22\\
	87	22\\
	88	22\\
	89	22\\
	90	22\\
	91	22\\
	92	22\\
	93	22\\
	94	22\\
	95	22\\
	96	22\\
	97	22\\
	98	22\\
	99	22\\
	100	22\\
	101	22\\
	102	22\\
	103	22\\
	104	22\\
	105	22\\
	106	22\\
	107	22\\
	108	22\\
	109	22\\
	110	22\\
	111	22\\
	112	22\\
	113	22\\
	114	22\\
	115	22\\
	116	22\\
	117	22\\
	118	22\\
	119	22\\
	120	22\\
	121	22\\
	122	22\\
	123	22\\
	124	22\\
	125	22\\
	126	22\\
	127	22\\
	128	22\\
	129	22\\
	130	22\\
	131	22\\
	132	22\\
	133	22\\
	134	22\\
	135	22\\
	136	22\\
	137	22\\
	138	22\\
	139	22\\
	140	22\\
	141	22\\
	142	22\\
	143	22\\
	144	22\\
	145	22\\
	146	22\\
	147	22\\
	148	22\\
	149	22\\
	150	22\\
	151	22\\
	152	22\\
	153	22\\
	154	22\\
	155	22\\
	156	22\\
	157	22\\
	158	22\\
	159	22\\
	160	22\\
	161	22\\
	162	22\\
	163	22\\
	164	22\\
	165	22\\
	166	22\\
	167	22\\
	168	22\\
	169	22\\
	170	22\\
	171	22\\
	172	22\\
	173	22\\
	174	22\\
	175	22\\
	176	22\\
	177	22\\
	178	22\\
	179	22\\
	180	22\\
	181	22\\
	182	22\\
	183	22\\
	184	22\\
	185	22\\
	186	22\\
	187	22\\
	188	22\\
	189	22\\
	190	22\\
	191	22\\
	192	22\\
	193	22\\
	194	22\\
	195	22\\
	196	22\\
	197	22\\
	198	22\\
	199	22\\
	200	22\\
	201	22\\
	202	22\\
	203	22\\
	204	22\\
	205	22\\
	206	22\\
	207	22\\
	208	22\\
	209	22\\
};
\addlegendentry{modified NCS setup}

\end{axis}
\end{tikzpicture}%
	}
	\caption{ Evolution of the bucket level with and without  terminal cost on the bucket level and with different prediction horizons.}
	\label{fig:evol_beta}
\end{figure}
\begin{figure}[!h]
	\centering
	\resizebox {0.5\textwidth} {!} {
		\input{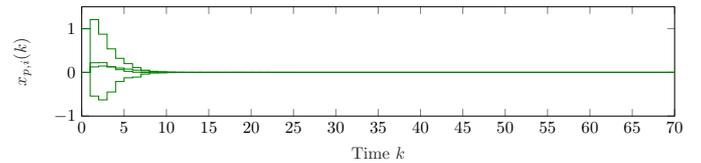}
	}
	\caption{ Evolution of the controlled plant state with terminal cost on the bucket level and $N=3$.}
	\label{fig:evol_plant_state}
\end{figure}
In the first example, we focus on the refilling behavior of the bucket. Figure \ref{fig:evol_beta} shows the evolution of bucket levels after the system was deflected by the initial condition. Notice first that when there is no terminal cost on the bucket level, the bucket will not refill to the upper sector, as already noticed in \cite{8723153}. This is because there is no term penalizing the bucket level in this setup. If a slight terminal cost with $\sigma = 10^{-6}$ is added in the setup of Section \ref{sec:problem_setup}, the bucket level converges to the upper sector after the plant state has converged as seen from Figures \ref{fig:evol_beta} and \ref{fig:evol_plant_state}.  We observe that the bound given in Theorem \ref{thm:upper_sector} is tight in this example. Note that for this setup, also a plot for $N=7$ is given, where the convergence interval is larger, as also expected from Theorem \ref{thm:upper_sector}. Lastly, we see that in the modified NCS setup from Section \ref{sec:periodic} with $\psi=9.93\cdot10^{-10}$ such that Assumption \ref{ass:sigma} is fulfilled, the bucket level converges exactly to the upper rim of the bucket due to the additional stage cost on the bucket level and the additional direct link in the network.

\begin{figure}[!h]
	\centering
	\resizebox {0.5\textwidth} {!} {
		\input{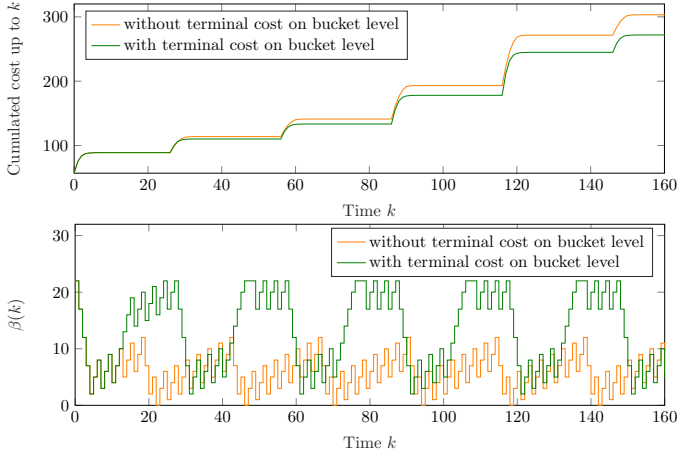}
	}
	\resizebox {0.5\textwidth} {!} {
%
%

%
\begin{tikzpicture}

\begin{axis}[%
width=4.844in,
height=1.5in,
at={(0.813in,0.378in)},
scale only axis,
xmin=0,
xmax=160,
xlabel style={font=\color{white!15!black}},
xlabel={Time $k$},
ymin=0,
ymax=32,
ylabel style={font=\color{white!15!black}},
ylabel={$\beta(k)$},
axis background/.style={fill=white},
legend style={legend cell align=left, align=left, draw=white!15!black}
]
\addplot[const plot, color=istorange] table[row sep=crcr] {%
0	22\\
1	17\\
2	12\\
3	7\\
4	2\\
5	5\\
6	8\\
7	3\\
8	6\\
9	9\\
10	4\\
11	7\\
12	10\\
13	5\\
14	8\\
15	11\\
16	6\\
17	9\\
18	12\\
19	7\\
20	2\\
21	5\\
22	0\\
23	3\\
24	6\\
25	1\\
26	4\\
27	7\\
28	2\\
29	5\\
30	8\\
31	3\\
32	6\\
33	9\\
34	4\\
35	7\\
36	10\\
37	5\\
38	8\\
39	11\\
40	6\\
41	9\\
42	12\\
43	7\\
44	2\\
45	5\\
46	0\\
47	3\\
48	6\\
49	1\\
50	4\\
51	7\\
52	2\\
53	5\\
54	8\\
55	3\\
56	6\\
57	9\\
58	4\\
59	7\\
60	10\\
61	5\\
62	8\\
63	11\\
64	6\\
65	9\\
66	12\\
67	7\\
68	2\\
69	5\\
70	0\\
71	3\\
72	6\\
73	1\\
74	4\\
75	7\\
76	2\\
77	5\\
78	8\\
79	3\\
80	6\\
81	9\\
82	4\\
83	7\\
84	10\\
85	5\\
86	8\\
87	11\\
88	6\\
89	9\\
90	12\\
91	7\\
92	2\\
93	5\\
94	0\\
95	3\\
96	6\\
97	1\\
98	4\\
99	7\\
100	2\\
101	5\\
102	8\\
103	3\\
104	6\\
105	9\\
106	4\\
107	7\\
108	10\\
109	5\\
110	8\\
111	11\\
112	6\\
113	9\\
114	12\\
115	7\\
116	2\\
117	5\\
118	0\\
119	3\\
120	6\\
121	1\\
122	4\\
123	7\\
124	2\\
125	5\\
126	8\\
127	3\\
128	6\\
129	9\\
130	4\\
131	7\\
132	10\\
133	5\\
134	8\\
135	11\\
136	6\\
137	9\\
138	12\\
139	7\\
140	2\\
141	5\\
142	0\\
143	3\\
144	6\\
145	1\\
146	4\\
147	7\\
148	2\\
149	5\\
150	8\\
151	3\\
152	6\\
153	9\\
154	4\\
155	7\\
156	10\\
157	5\\
158	8\\
159	11\\
160	6\\
161	9\\
162	12\\
163	7\\
164	2\\
165	5\\
166	0\\
167	3\\
168	6\\
169	1\\
170	4\\
171	7\\
172	2\\
173	5\\
174	8\\
175	3\\
176	6\\
177	9\\
178	4\\
179	7\\
180	10\\
181	5\\
182	8\\
183	11\\
184	6\\
185	9\\
186	12\\
187	7\\
188	2\\
189	5\\
190	0\\
191	3\\
192	6\\
193	1\\
194	4\\
195	7\\
196	2\\
197	5\\
198	8\\
199	3\\
200	6\\
201	9\\
202	4\\
203	7\\
204	10\\
205	5\\
206	8\\
207	11\\
208	6\\
209	9\\
};
\addlegendentry{without terminal cost on bucket level}

\addplot[const plot, color=istgreen] table[row sep=crcr] {%
0	22\\
1	17\\
2	12\\
3	7\\
4	2\\
5	5\\
6	8\\
7	3\\
8	6\\
9	9\\
10	4\\
11	7\\
12	10\\
13	13\\
14	16\\
15	19\\
16	14\\
17	17\\
18	20\\
19	15\\
20	18\\
21	21\\
22	16\\
23	19\\
24	22\\
25	17\\
26	20\\
27	22\\
28	17\\
29	12\\
30	7\\
31	2\\
32	5\\
33	8\\
34	3\\
35	6\\
36	9\\
37	4\\
38	7\\
39	10\\
40	5\\
41	8\\
42	11\\
43	14\\
44	17\\
45	20\\
46	22\\
47	22\\
48	22\\
49	17\\
50	20\\
51	22\\
52	17\\
53	20\\
54	22\\
55	17\\
56	20\\
57	22\\
58	17\\
59	12\\
60	7\\
61	2\\
62	5\\
63	8\\
64	3\\
65	6\\
66	9\\
67	4\\
68	7\\
69	10\\
70	5\\
71	8\\
72	11\\
73	14\\
74	17\\
75	20\\
76	22\\
77	22\\
78	22\\
79	17\\
80	20\\
81	22\\
82	17\\
83	20\\
84	22\\
85	17\\
86	20\\
87	22\\
88	17\\
89	12\\
90	7\\
91	2\\
92	5\\
93	8\\
94	3\\
95	6\\
96	9\\
97	4\\
98	7\\
99	10\\
100	5\\
101	8\\
102	11\\
103	14\\
104	17\\
105	20\\
106	22\\
107	22\\
108	22\\
109	17\\
110	20\\
111	22\\
112	17\\
113	20\\
114	22\\
115	17\\
116	20\\
117	22\\
118	17\\
119	12\\
120	7\\
121	2\\
122	5\\
123	8\\
124	3\\
125	6\\
126	9\\
127	4\\
128	7\\
129	10\\
130	5\\
131	8\\
132	11\\
133	14\\
134	17\\
135	20\\
136	22\\
137	22\\
138	22\\
139	17\\
140	20\\
141	22\\
142	17\\
143	20\\
144	22\\
145	17\\
146	20\\
147	22\\
148	17\\
149	12\\
150	7\\
151	2\\
152	5\\
153	8\\
154	3\\
155	6\\
156	9\\
157	4\\
158	7\\
159	10\\
160	5\\
161	8\\
162	11\\
163	14\\
164	17\\
165	20\\
166	22\\
167	22\\
168	22\\
169	17\\
170	20\\
171	22\\
172	17\\
173	20\\
174	22\\
175	17\\
176	20\\
177	22\\
178	17\\
179	20\\
180	22\\
181	17\\
182	20\\
183	22\\
184	17\\
185	20\\
186	22\\
187	17\\
188	20\\
189	22\\
190	17\\
191	20\\
192	22\\
193	17\\
194	20\\
195	22\\
196	17\\
197	20\\
198	22\\
199	17\\
200	20\\
201	22\\
202	17\\
203	20\\
204	22\\
205	17\\
206	20\\
207	22\\
208	17\\
209	20\\
};
\addlegendentry{with terminal cost on bucket level}

\end{axis}
\end{tikzpicture}%
	}
	\caption{ Comparison of cumulated cost (top) and  evolution of the bucket level (bottom) with and without terminal cost on the bucket level.}
	\label{fig:comp_cost}
\end{figure}
In the second example, we compare the performance of the control with and without terminal cost on the bucket. For this purpose, we consider six set point changes and cumulate the resulting stage cost, which does not pose a problem in the considered unconstrained setup. We consider only the setup from Section \ref{sec:problem_setup} and leave out the modified NCS setup from Section \ref{sec:periodic}. This is because more possible transmissions are overall available in the latter due to the additional direct link. Figure \ref{fig:comp_cost} shows the cumulated stage cost under the rollout controller up to time $k$ and the resulting evolution of the bucket level for the NCS setup from Section \ref{sec:problem_setup}. The cumulated costs after reaching the first set point are equal by visual comparison although in theory, the setup with terminal cost should have a slightly larger cumulated cost. After the second set point change, the cumulated cost with terminal cost on the bucket level is clearly lower than the cost without.  With every set point change, the difference becomes even larger. We observe from Figure \ref{fig:comp_cost} (bottom) why this is the case: By adding a slight terminal cost on the bucket level, the bucket refills to the upper sector after each set point change. This results in more possible transmissions being available at the upcoming set point change, which in turn leads to faster convergence and hence to a lower cost.
\section{Summary and Outlook}
\label{sec:conclusion}
In this paper, we considered a token bucket TS acting as a traffic shaper for the communication of control updates over a shared network. Our main contribution was to develop mechanisms under which tokens are saved once the plant state has converged. By this measure, the controller may react appropriately to unforeseen operating conditions, since the tokens were able to recover during the uncritical phases. We considered two different setups. In the first, control was performed over a network where all transmissions needed to fulfill the token bucket TS. We achieved convergence of the bucket level to the upper sector by adding an arbitrarily small terminal cost on the bucket level. Second, we demonstrated that by extending this network by a direct link and adding a small stage cost on the bucket level, the bucket level converges exactly to the upper rim. The numerical example showed that the proposed mechanisms are effective and revealed that they may significantly improve control performance when several set point changes occur successively.

We conjecture that similar results as presented in Section \ref{sec:extended_term} could be obtained without the restriction that $\frac{c}{g}\notin\mathbb{I}$, since convergence is indeed observed in simulations also for this case. To provide a theoretical guarantee could be the topic of future work.


\bibliography{ifacconf}             
                                                   







\end{document}